\documentclass{sig-alternate}

\usepackage{etoolbox}
\makeatletter
\patchcmd{\maketitle}{\@copyrightspace}{}{}{}
\makeatother

\usepackage{booktabs}

\usepackage{flushend}

\begin{document}
%

\title{Video (GIF) Sentiment Analysis Using Large Scale Auto Generated Mid-Level Ontology}

%
%
%
%
%

\numberofauthors{3} 
%
\author{
%
%
\alignauthor
Cai Zheng\\
       \affaddr{Xiamen University}\\
       \email{31520131153288\\@stu.xmu.edu.cn}
\alignauthor
Cao Donglin\\
       \affaddr{Xiamen University}\\
       \email{another@xmu.edu.cn}
\alignauthor
Ji Rongrong\\
       \affaddr{Xiamen University}\\
       \email{jirongrong@xmu.edu.cn}
}


\maketitle

\begin{abstract}
With faster connection speed, Internet users are now making social network a huge reservoir of texts, images and video clips (GIF). Sentiment analysis for such online platform can be used to predict political elections, evaluates economic indicators and so on. However, GIF sentiment analysis is quite challenging, not only because it hinges on spatio-temporal visual contentabstraction, but also for the relationship between such abstraction and final sentiment remains unknown.In this paper, we dedicated to find outsuch relationship.We proposed a SentiPairSequence basedspatio-temporal visual sentiment ontology, which forms the mid-level representations for GIFsentiment. The establishment process of SentiPair contains two steps. First, we construct the Synset Forest to define the semantic tree structure of visual sentiment label elements. Then, through theSynset Forest, we organically select and combine sentiment label elements to form a mid-level visual sentiment representation. Our experiments indicate that SentiPair outperforms other competing mid-level attributes. Using SentiPair, our analysis frameworkcan achieve satisfying prediction accuracy (72.6\%). We also opened ourdataset (GSO-2015) to the research community. GSO-2015 contains more than 6,000 manually annotated GIFs out of more than 40,000 candidates. Each is labeled with both sentiment and SentiPair Sequence.
\end{abstract}




\section{INTRODUCTION}
Nowadays, Internet users are making every social network (such as \textit{Facebook, Twitter, Weibo, }et al.) a huge reservoir of text, images and video clips (GIF). Men and women are sharing their lives and opinions at online platforms, which gives a new direction for applications such as political election prediction, economic indicator measurement, policy feedback evaluation and so on.\\

With faster Internet connection, people are more willing to post GIF videos other than static images. According to a recent study\cite{Yuan_Sentribute} the total proportion of visual content from all shared links on Twitter is 36\%. However, Industries deviate from user habit by taking social network sentiment analysis as social network text analysis. Research for GIF sentiment analysis is still in its infancy. This is partly because the inherent features of online GIF videos. GIF videos spreading in online social networks share some common features.\\

\begin{figure}
\centering
\epsfig{file=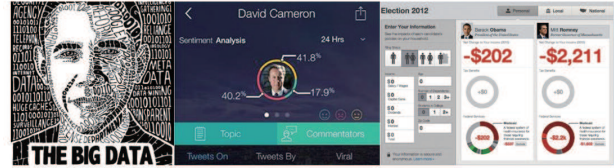, width=3.25in}
\caption{Sentiment analysis behinds political election}
\end{figure}

\textbf{Noise}: Junk information can interfere the judgment of GIF sentiment. We studied the composition of GIF videos. Among 6,000 GIF videos, we found around 71.55\% of them are mixed with other forms of information. The most common (around 34.49\%) noise is explanative texts. They can reverse the GIF sentiment. Other noises include motion blur, unexpected illumination change and so on.\\

\textbf{Irregular Length}:Online GIF video¡¯s length is irregular. The average length of 40,000 GIF videos we collected is 17.82s, with the longest GIF video (3¡¯13¡¯¡¯22) and the shortest GIF video (0¡¯0¡¯¡¯3¡¯).The irregular length makes it difficult to analyze.\\

\textbf{Symbolism}: Symbolism refers to the fact that final sentiment judgment doesn't rely on object appeared inthe GIF, but the hidden message that object symbolized. For example, Sylvester Stallone symbolizes brave, perhaps in a video Stallone shows up, but it is brave that influence the final sentiment judgment.\\

\textbf{Abstraction}: GIF sentiment hinges on high-level abstractions. Take the GIF video showed in Figure 4 as an example, GIF video¡¯s sentiment changes with the variation of the girl¡¯s facial expression. In general, to utilize the abstractions, there are several problems to solve.\\

\begin{enumerate}
  \item How to describe abstractions: A mid-level representation system has to be designed.
  \item How to detect abstractions: Given that representation system, we should be able to detect them.
  \item How to get sentiment from abstractions: The relationship between abstractions and the final sentiment judgment should be modeled.
\end{enumerate}

\begin{figure}
\centering
\epsfig{file=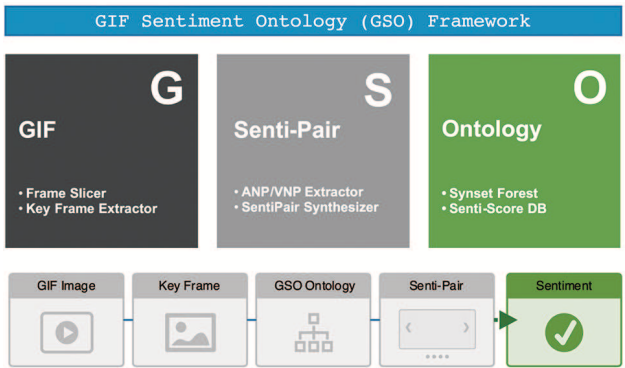, width=3.25in}
\caption{The architecture of GSO framework}
\end{figure}

In a word, the problem of GIF sentiment analysis is complex and challenging. But we consider it as a process of two steps. The first step is to extract mid-level representations for those abstractions (\emph{Smile Face}, \emph{Body Movement}, et al. We call them SentiPairs). Then we get SentiPair Sequence by combing SentiPairs in the order of their occurrence. In the second step, using the model we built, a sentiment judgment is made based on the SentiPair Sequence.
Restricted to the accuracy of object detection,object tracking, et al. It is not easy to track the abstractions simultaneously, but we looked from another angle.\\

By generating a collection of mid-level representations, modeling the relationship between SentiPair and sentiment, and opening the first procedure to other fast-developing research fields (video classification, et al.), we can get GIF video¡¯s sentiment.To that end, we address in this work two major challenges:

\begin{enumerate}
  \item To find out a common and effective representation for GIF videos.
  \item Model the relationship between representations and the final sentiment judgment.
\end{enumerate}

In particular, we make the following contributions:
\begin{description}
  \item[a] We proposed a new mid-level representation: \texttt{SentiPair Sequence.} SentiPair Sequenceis based on semantic tree structure of visual sentiment label elements¡ªthe Synset Forest. With SentiPair Sequence and Synset Forest, we built the GIF Sentiment Ontology (GSO). Our experiments indicate that SentiPairoutperforms other mid-level attributes and our framework can achieve satisfying detect accuracy.
  \item[b] We built a large manually labeled GIF sentiment dataset (GSO-2015 dataset) from 40,000+ GIF videos spread around social networks. This dataset contains not only GIF picture¡¯s sentiment but also the SentiPair Sequence. GSO-2015 is free and will be open tothe public to promote further research on visual sentiment.
\end{description}

\section{RELATED WORK}
Different researchers have developed different systems. Recent studies lay their emphasis on subject-free image sentiment analysis, subject-specified video sentiment analysis and subject-free video sentiment analysis.\\

\begin{figure}
\centering
\epsfig{file=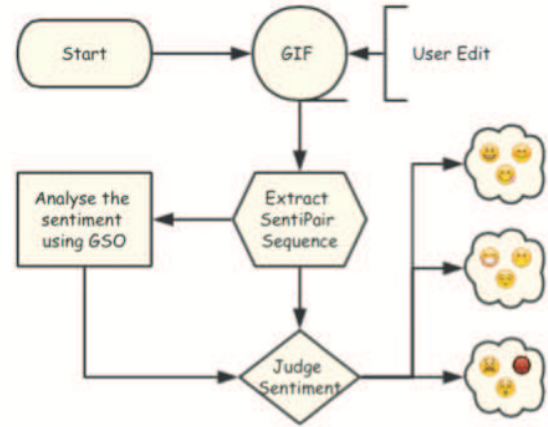, width=3.25in}
\caption{Working flow using GSO framework}
\end{figure}

For subject-free image sentiment analysis, \cite{Quanzeng_PCNN} used the Progressive CNN network and bypassed the mid-level features. However, the number of neurons and connections is huge due to the \textbf{abstract} nature of visual sentiment. Deep network needs vast amount of less \textbf{noisy} labeled training instances to adjust the equal vast amount of neurons. Otherwise, it will get stuck into local optimum. Both \cite{Borth_VSO} and \cite{Yuan_Sentribute} proposed to employ mid-level entities or attributes as features for subject-free image sentiment analysis.In \cite{Borth_VSO}, 1200 adjective noun pairs (ANP), which may correspond to different levels of different emotions, are extracted. These ANPs are used as queries to crawl images from Flickr. Next, pixel-level features of images in each ANP are employed to train 1200 ANP detectors. The responses of these 1200 classifiers can then be considered as mid-level features for visual sentiment analysis. The work in \cite{Yuan_Sentribute} employed a similar mechanism. The main difference is that 102 scene attributes are used instead.\\

For subject-specified video sentiment analysis, \cite{Morency} proposed a framework utilizes video sound and facial expression to analyze \emph{interview clips}. They focused on the sentiment analysis towards video with fixed content, similar pattern and average noise. The experiment result is promising, but due to the subject is specified, the method can¡¯t be used to deal with large-scale social network¡¯s GIF videos.\\

For subject-free video sentiment analysis, \cite{GIFGIF} proposed to use the features such as color histogram to train a framework for online GIF sentiment analysis. They did proposed good GIF emotion dataset. But since the problem of subject-free video sentiment analysis hinges on some unsolved problems, namely object classification, facial expression recognition, et al\., we can't rely on a unified framework for such an \textbf{Abstract} problem.\\

\begin{figure}
\centering
\epsfig{file=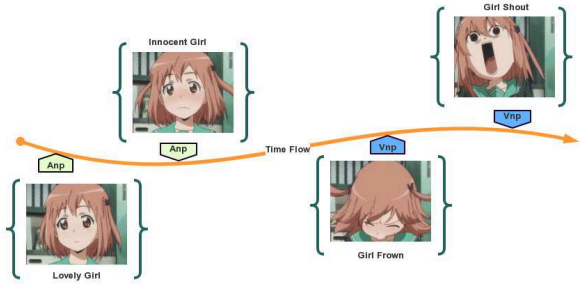, width=3.25in}
\caption{SentiPair Sequence}
\end{figure}

\section{GSO FRAMEWORK}
We dedicated to find out the essential relationship between abstractions and sentiment for GIF videos.To achieve that goal, we developed an auto-generated ontology based on vast amount of user-generated content. We also developed GSO Framework to implement the ontology. The architecture of the GSO Framework we employ is shown in Figure 2. The details of the proposed framework will be described in the following sections.\\

\subsection{SentiPair Sequence}
When considering the problem of GIF sentiment analysis. We should firstly figure out a way to represent a GIF video. A good representation should follow several criterions:

\begin{enumerate}
  \item \textbf{Descriptive}: should be able to describe the abstractions.
  \item \textbf{Detective}: should be able to be detected easily.
  \item \textbf{Easy}: should be able to be understood.
  \item \textbf{Flexible}: should be able to be extended.
\end{enumerate}

To resolve these criterions, we introduced the SentiPair Sequence. A SentiPair Sequence is a sequence of SentiPairs, while the SentiPair is the joint name of Adjective Noun Pair (ANP) and Verb Noun Pair (VNP).Each SentiPair refers to either a concrete concept like \emph{smile face} or a specific motion like \emph{falling cup}. In a SentiPair Sequence, SentiPairs are sorted by the order of their occurrence.\\

Figure 4 shows a typical SentiPair Sequence.As we can see, the girl in the video acts differently. At the very first, The girl is smiling and hence the first SentiPair indicates \emph{Lovely Girl}, In the next frame, the girl looked a bit worried, and the second SentiPair is \emph{Innocent Girl}, With the third SentiPair indicates \emph{Girl Frown}, we can find out that the girl looks sad, which contains a negative sentiment tendency. In the last frame, the girl failed to suppress her feeling and the SentiPair indicates \emph{Girl Shout}. As a result, we can conclude the SentiPair Sequence of this GIF video as: \emph{Lovely Girl}, \emph{Innocent Girl}, \emph{Girl Frown}, and \emph{Girl Shout}.\\

SentiPair Sequence describes the concepts associated with sentiment judgment. In general, SentiPair Sequence carries two kinds of concepts. The first one is the existed object (by ANP), and the second one is object¡¯s movement (by VNP).\\

\begin{figure}
\centering
\epsfig{file=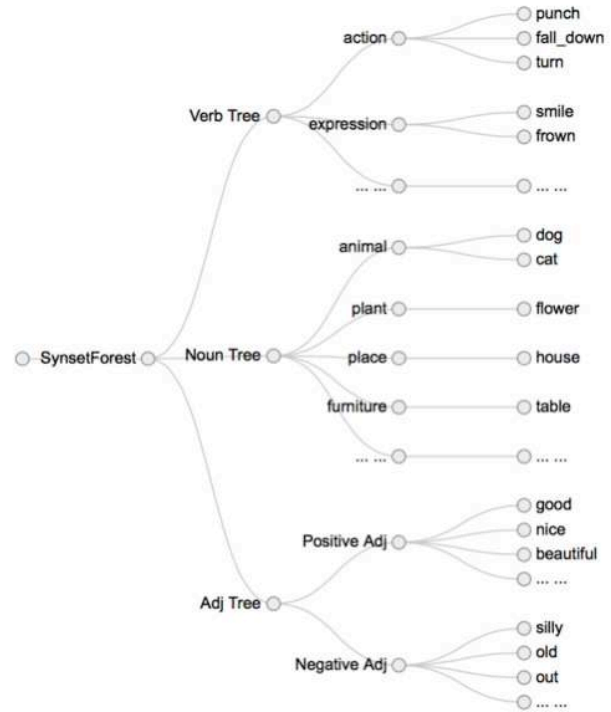, width=3.25in}
\caption{An overlook of the Synset Forest}
\end{figure}

\begin{figure*}
\centering
\epsfig{file=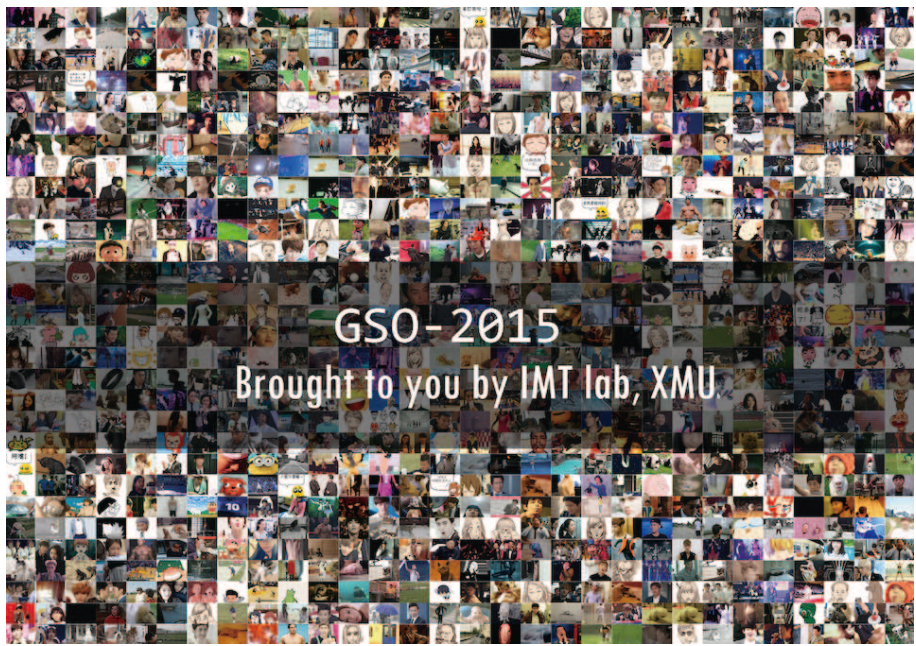, width=4.75in}
\caption{The GSO-2015 Dataset}
\end{figure*}

\textbf{Adjective Noun Pairs(ANP)} was firstly introduced by \cite{Borth_VSO}, compared to nouns or adjectives only, an ANP can turn a neutral noun like \emph{dog} into an ANP with strong sentiment like \emph{cute dog} by adding an adjective with a strong sentiment. And the combined phrases also make the concepts more detectable than adjectives (like \emph{beautiful}), which are too abstract.\cite{Borth_VSO}\\

Inspired by previous work \cite{Borth_VSO}, the combined phrase of ANP has a better detect accuracy than single word, easier to be interpreted, and the whole structure is portable and flexible because we can add/remove ANPs dynamically. \\

However, our experiment shows that ANP itself alone is not enough in the field of video sentiment analysis. We introduced \textbf{Verb Noun Pair(VNP)}. VNP shares the same structure of ANP, accordingly, shares the features of portable, easy to be detected and interpreted.\\

\subsection{Synset Forest}
SentiPairs are built on three kinds of words: adjectives, verbs and nouns. In order to build ANP/VNP, we should first build the collection of words to choose from. We concluded some criterions for a good word collection.

  \begin{table}[htbp]
 \caption{\label{tab:tab1}Accuracy without attribute selection}
  \centering
 \begin{tabular}{ccccc}

  \toprule
Algorithm & Prec. & Recall & FScore & Acc. \\

  \midrule
 Navie Bayes & 0.697 & 0.704 & 0.686 & 0.706\\
  SMO & 0.791 & 0.727 &	0.719 &	0.726\\
 Logistic &	0.705 &	0.704 &	0.704 &	0.703\\
AdaBoost &	0.668 &	0.629 &	0.528 &	0.629\\
Rand.Forest &	0.713 &	0.715 &	0.711 &	0.714\\
  \bottomrule
 \end{tabular}
\end{table}

  \begin{table}[htbp]
 \caption{\label{tab:tab2}Accuracy with Correlation Based Subset}
  \centering
 \begin{tabular}{ccccc}

  \toprule
Algorithm & Prec. & Recall & FScore & Acc. \\

  \midrule
Navie Bayes &	0.708	& 0.704 &	0.675 &	0.704\\
SMO &	0.722 &	0.72 &	0.697 &	0.719\\
Logistic &	0.731 &	0.727 &	0.703 &	0.726\\
AdaBoost	& 0.668	& 0.629 &	0.528 &	0.629\\
Rand.Forest &	0.729	& 0.724 &	0.701 &	0.724\\

  \bottomrule
 \end{tabular}
\end{table}

\begin{enumerate}
  \item \textbf{Coverage}, a good word collection should cover as much domains as possible in order to convey the information
  \item \textbf{Discrepancy}, words of similar meanings should appear only once to prevent ambiguous ANP/VNP.

  \item \textbf{Sentiment relation}, all the words in the collection should have a clear sentiment relation.
\end{enumerate}

We introduced \textbf{Synset Forest} to resolve these three criterions. The Synset Forest is a forest consists of three trees, namely adjective tree, verb tree and the noun tree. An overlook of all three trees can be found at Figure 5.\\

In the \textbf{Wordnet}\cite{wordnet}, Synsets are interlinked by means of conceptual-semantic and lexical relations.By proposing the Synset Forest, we modeled a unified semantic and concept architecture. The Synset Forest acts as a collection of candidate words for Adjective Noun Pairs and Verb Noun Pairs. Since each node comes with a sentiment score, the weight for each ANP/VNP is decided at the first place.

\subsection{GSO-2015 Dataset}
We built a new GIF video dataset from one of the most popular micro-blog provider. All the GIF videos were posted by online users and were collected automatically. We built 40,000+ distinct candidates. These candidates were then manually labeled in the fashion of GIF Sentiment Ontology. This work is possible owing to the crowd intelligence. We recruited 7 workers. Each worker was shown one GIF video and was expected to accomplish two \emph{tasks}. \textbf{Task 1} is to depict the given GIF using SentiPair Sequence. To be more specific, for each GIF, SentiPairs were chosen by the worker. And each SentiPair consists either of an adjective and a noun(ANP) or a verb and a noun (VNP). Figure 4 illustrates the flow of SentiPairs and the corresponding GIF. In \textbf{Task 2} workers were expected to give the image an overall sentiment judgment(Positive/Negative/Neutral/Can't Judge).

\section{EXPERIMENTS}
We designed two experiments. The first experiment is designed to evaluate the performance of SentiPair Sequences we proposed. Different models applied and the evaluation metric is the accuracy. Moreover, we explored the possibility to simplify the problem by introducing feature selection.\\

  \begin{table}[htbp]
 \caption{\label{tab:tab3}Accuracy with Correlation Based Subset}
  \centering
 \begin{tabular}{cccc}

  \toprule
Type &	SMO &	Logistic &	Rand. Forest \\

  \midrule
ANP only &	0.702 &	0.696 &	0.696\\
VNP only &	0.658 &	0.662 &	0.655\\
SentiPair &	0.726 &	0.703 &	0.714\\

  \bottomrule
 \end{tabular}
\end{table}

The second experiment is designed to compare the performance of our framework to the state-of-the-art representation VSO \cite{Borth_VSO}. The evaluation metric is the accuracy as well.\\

We choose to use the GSO-2015 dataset to train the sentiment classifiers. One of the advantages of GSO is its ability to convey temporal information(through SentiPair Combination). The training set consists of 1124 positive instances (60.3\%), 146 negative instances (7.8\%) and 599 neutral ones (32.1\%).

\subsection{Baseline}
We compare the performance of GIF Sentiment Ontology(GSO) on GSO-2015 dataset with the state-of-the-art image sentiment analysis framework the VSO\cite{Borth_VSO}.

\subsection{GIF Sentiment Analysis With Bag of Word}
We decided to explore a simple way to get sentiment from SentiPairs. A straightforward way is BoW. Table 1 shows the performance of BoW using different classifiers. We find the Logistic Regression and SMO are the best choices (the same as \cite{Borth_VSO}).On the other hand, each algorithm is able to correctly classify a large proportion of the testing instances(around 70\%). In other words, classifier is not that important if we have SentiPair Sequences.\\

We also explored the possibility to simplify the problem by introducing feature selection. We used Correlation Based Subset (implemented by WEKA\cite{weka}) and the results are shown in Table 2. We find that feature selection can provide similar detect accuracy while using less mid-level attributes.

\subsection{SentiPair V.S. ANP}
To evaluate the mid-level representation we proposed, we compare the performance of SentiPair Sequence with both single ANP and single VNP. The numbers in Table 3 is the accuracy indicating the instances been correctly classified. The result indicated that SentiPair Sequence outperformed ANP. We believe this is because GIF sentiment is affected by motion information, which is expressed by VNP.

\section{CONCLUSION}
In this paper, we find out the essential relationship between abstractions and sentiment for GIF videos. As a mid-level representation for GIF videos, SentiPair Sequence was proposed. SentiPair is built on Synset Forest, while the Forest consists of 3 trees of Synset. Each Synset is a word with a specific sense.  Our experiments suggest that SentiPair Sequence can abstract video¡¯s spatio-temporal information and it outperforms competing representations such as ANP or VNP. The ontology is automatically established and the experiments suggest that the prediction accuracy is 72.6\%. We leave the first process (such as object detection) to the relevant researchers because we believe this process can be done simultaneously. Every single step made in the first process will improve the overall performance. We also opened our dataset GSO-2015 to the public. GSO-2015 contains more than 6,000 manually annotated GIF videos selected from more than 40,000 candidates. Each video is labeled with both sentiment and SentiPair Sequence. We believe it will be helpful for further researchers.



%
\bibliographystyle{unsrt}

\begin{thebibliography}{1}

\bibitem{Yuan_Sentribute}
You~Quanzeng Yuan~Jianbo, Mcdonough~Sean and Luo Jiebo.
\newblock Sentribute: Image sentiment analysis from a mid-level perspective.
\newblock In {\em Proceedings of the Second International Workshop on Issues of
  Sentiment Discovery and Opinion Mining}, WISDOM '13, pages 10:1--10:8, New
  York, NY, USA, 2013. ACM.

\bibitem{Quanzeng_PCNN}
Quanzeng You, Jiebo Luo, Hailin Jin, and Jianchao Yang.
\newblock Robust image sentiment analysis using progressively trained and
  domain transferred deep networks.
\newblock In {\em Robust Image Sentiment Analysis Using Progressively Trained
  and Domain Transferred Deep Networks}, 2015.

\bibitem{Borth_VSO}
Damian Borth, Rongrong Ji, Tao Chen, Thomas Breuel, and Shih-Fu Chang.
\newblock Large-scale visual sentiment ontology and detectors using adjective
  noun pairs.
\newblock In {\em Proceedings of the 21st ACM International Conference on
  Multimedia}, MM '13, pages 223--232, New York, NY, USA, 2013. ACM.

\bibitem{Morency}
Louis-Philippe Morency, Rada Mihalcea, and Payal Doshi.
\newblock Towards multimodal sentiment analysis: Harvesting opinions from the
  web.
\newblock In {\em Proceedings of the 13th International Conference on
  Multimodal Interfaces}, ICMI '11, pages 169--176, New York, NY, USA, 2011.
  ACM.

\bibitem{GIFGIF}
Brendan Jou, Subhabrata Bhattacharya, and Shih-Fu Chang.
\newblock Predicting viewer perceived emotions in animated gifs.
\newblock In {\em Proceedings of the ACM International Conference on
  Multimedia}, MM '14, pages 213--216, New York, NY, USA, 2014. ACM.

\bibitem{wordnet}
George~A. Miller.
\newblock Wordnet: A lexical database for english.
\newblock In {\em Communications of the ACM}, volume 38,No.11, pages 39--41,
  1995.

\bibitem{weka}
Antti Puurula and Sung-Hyon Myaeng.
\newblock Integrated instance- and class-based generative modeling for text
  classification.
\newblock In {\em Proc 18th Australasian Document Computing Symposium}, pages
  66--73. ACM, 2013.

\end{thebibliography}

%
%


\end{document}